# TURBULENCE KINETIC ENERGY DISTRIBUTION AND HEAT TRANSFER IN A POROUS LAYER INDUCED BY BLUFF BODY VORTEX SHEDDING

**Thibaut K Kemayo[1], Justin Courter[1], Vishal Srikanth[1], Chadwick Jetti[1], Rodrigo R Caballero[1], Andrey V Kuznetsov[1*]**

[1]Department of Mechanical and Aerospace Engineering, North Carolina State University
Raleigh, NC 27695-7910, USA

## ABSTRACT

When a turbulent vortex impinges on a porous layer, it creates a complex multiscale interaction: the wake structures that form in the free fluid engage with the intricate geometry of the pores, and this interplay governs both the turbulent energy budget and the rate of heat transfer. Here we use interface-resolved two-dimensional direct numerical simulations (DNS) to examine how a bluff-body wake impinges on an in-line porous array heated to maintain a constant wall temperature. The Reynolds number is fixed at $Re = 10000$, and the porosity is varied between $\phi = 0.80$ and $\phi = 0.95$. In all cases, the incoming von Kármán vortices undergo rapid breakdown at the porous/fluid interface and do not persist as coherent macroscale structures within the porous layer. The interface instead acts as a spectral filter: large-scale wake energy is strongly attenuated, while turbulence is regenerated locally within the matrix via shear layers and microscale vortex shedding around individual obstacles. Thermal statistics show that the lower-porosity medium produces higher local and surface-averaged Nusselt numbers across representative interface and interior locations. This is consistent with the stronger shear and enhanced fluid/solid thermal interaction associated with the larger surface-area-to-volume ratio. These results clarify the mechanisms by which wake-driven turbulence is converted into pore-scale motions and how porosity tunes the balance between turbulence attenuation and convective heat transfer in porous coatings and inserts.

## 1. INTRODUCTION

Turbulence kinetic energy (TKE) distribution and convective heat transfer in porous layers impacted by bluff-body vortex shedding form a multiscale problem that links wake dynamics in the clear fluid, interfacial turbulence production at the porous/fluid boundary, and pore-scale turbulence dissipation inside the matrix. Porous layers and coatings are widely used to manage momentum and heat in compact heat exchangers, catalytic beds, porous aerodynamic surfaces, and geophysical canopies [1–4]. In many of these systems, a turbulence-generating element upstream, such as a bluff body or wind gust, or turbulence generated by a cooling fan creates an unsteady wake whose vortices strike a porous layer downstream. These vortices inject large-scale turbulent motions at the interface that must be filtered by the geometry and redistributed toward pore-scale motions before they can efficiently enhance heat removal within the medium.

Despite extensive research work on turbulence in porous media, the pathways by which macro-scale turbulence kinetic energy enters a porous/fluid interface and is transported and dissipated inside a porous array remain only partially resolved. Interface-resolved simulations suggest that the interfacial TKE budget is controlled by a competition among production, pressure–strain redistribution, and turbulent diffusion. Simulations also suggest that these terms are highly sensitive to permeability, porosity, and array geometry [1]. Wrapping a bluff body in a porous layer alters the shedding frequency, the organization of the wake, and the near-wake spectra, sometimes stabilizing certain modes and sometimes delaying shedding without suppressing it completely. Consequently, both the intensity of large-scale wake structures and their ability to penetrate downstream into a porous layer depend strongly on the properties of the porous matrix itself [2]. Moreover, recent numerical studies of turbulent jet impingement interacting with engineered porous lattice structures indicate that the geometry and characteristic length scales of the porous matrix can significantly enhance local heat transfer while modifying flow organization and pressure losses, highlighting the strong coupling between turbulence, pore-scale structure, and thermal performance [5]. Inside the porous region, the accessible range of turbulent scales is constrained by the pore geometry: energetic externally generated eddies are broken down as they encounter the solid skeleton, but the resulting fluctuations can still drive interfacial production and localized dissipation along the solid surfaces, where fluid–solid interactions enhance heat transfer. In this context, high-fidelity direct numerical simulations based on high-order spectral difference methods with immersed boundary treatments have recently been shown to accurately resolve multiscale turbulence–porous interactions in complex geometries, providing detailed insight into both wake penetration and pore-scale dynamics [6].

*Corresponding Author: avkuznet@ncsu.edu

This interplay creates a design trade-off: flow-control strategies that weaken large-scale unsteadiness may either degrade or improve thermal performance, depending on how effectively the injected energy is redirected into pore-scale motions that exchange heat with the solid matrix. Numerical and experimental studies on porous-coated cylinders and porous inserts report heat-transfer enhancements that are strongly correlated with permeability, coating thickness, and the associated form drag, thus motivating performance metrics that explicitly balance Nusselt-number gains against the additional pressure drop [3,4]. There is, however, a need for simplified configurations that directly connect bluff-body vortex shedding, interfacial TKE redistribution, and heat transfer in a resolved porous layer while remaining inexpensive enough to allow studying the effects of systematic parameter variations.

In this study, we address that gap using a two-dimensional canonical configuration: the wake of a bluff body impinging on a downstream porous layer idealized as an in-line array of obstacles. We solve the incompressible, unsteady 2D Navier–Stokes equations with high spatial resolution such that the dominant shedding dynamics and interfacial production are resolved. This 2D framework isolates the essential coupling between large-scale wake structures in the clear fluid and the geometry-filtered motions inside the porous layer while enabling systematic variation of porosity. Although genuinely three-dimensional effects (such as spanwise instabilities and out-of-plane pore jets) are not resolved, the present approach is well suited to answer the broader question of whether macroscale turbulence can persist inside the porous layer and to understand how turbulence transport occurs at the interface.

Beyond reporting global metrics such as the pressure drag and Nusselt number, we also examine the microscale turbulence kinetic budget considering individual components from production to dissipation rate. Our objectives are to: (i) delineate interaction regimes between external large-scale wake structures and the porous layer; (ii) identify the scale-interaction mechanisms that govern penetration and breakdown; and (iii) establish how turbulence impingement influences convective heat transfer. Our numerical simulations revealed that the macroscale turbulent structure is near-instantaneously broken down at the porous/fluid interface. We report the mechanism of this breakdown followed by the impact on heat transfer both at the porous/fluid interface and deep inside the porous layer where periodic flow patterns are expected to emerge.

## 2. METHODOLOGY

**2.1 Computational Geometry.** The simulation domain (figure 1) comprises a square macroscale solid obstacle of side $D$ placed in a uniform free stream, followed by a porous layer modeled as a two-dimensional inline array of microscale square solid obstacles. The solid obstacles in the porous layer have a characteristic size $s$ and the distance between the centers of the solid obstacles is $d$. We vary the porosity $\phi = \{0.80, 0.95\}$ by varying $s$ at fixed $d$ to preserve the scale separation between the macroscale solid obstacle and the porous medium.

As illustrated in figure 2, the computational domain is further subdivided for the analysis into free-flow regions above and below the macroscale obstacle, an impingement region where the vortex generated by the macroscale obstacle directly interacts with the porous layer, and porous subregions classified as interface, interior, and exterior. The porosity of the porous medium is calculated as:

$$\phi = 1 - \left(\frac{s}{d}\right)^2 \tag{1}$$

The boundary conditions used in the simulation are as follows. The inlet is treated as a constant velocity inlet of 1 m/s, and the outlet is a constant pressure boundary specifying a gauge pressure of 0. Solid walls are treated as no-slip boundaries. The top and bottom boundaries are treated as periodic boundaries. The Reynolds number of the flow is specified as 10000, and is calculated as:

$$R_e = \frac{\rho u_{in} d}{\mu} \tag{2}$$

where $\rho$ is the fluid density (1 kg $m^{-3}$) and $\mu$ is the fluid viscosity (0.0001 Pa s). The thermal boundary conditions consider that the porous obstacles are maintained at a constant wall temperature of 350 K. The inlet temperature and the wall temperature of the macroscale solid obstacles are 300 K. The specific heat of the fluid is 7000 J $kg^{-1}$ $K^{-1}$ and the thermal conductivity is 1 W $m^{-1}$ $K^{-1}$.

In order to resolve the microscale turbulence structures at a feasible computational cost, the geometry is assumed to be 2D, similar to prior studies that employed 2D simulation models to resolve turbulence and develop fundamental scaling behaviors [7–10]. The objective of this study is to develop a theoretical, qualitative understanding of how macroscale vortices interact with the microscale pores of a porous layer. To this end, we employ the 2D approximation as a theoretical limit, isolating the dynamics of coherent structures from stochastic background turbulence. We acknowledge that 2D DNS inherently neglects the vortex stretching mechanism essential for the forward energy cascade in 3D flows. While this omission prevents the prediction of realistic vortex breakdown, it enables a rigorous characterization of the fundamental shedding modes and their interaction with pore boundaries.

Consequently, we anticipate a prolonged survival of large-scale eddies in 2D turbulence compared to 3D. Precise quantitative predictions of drag or Nusselt numbers are outside the scope of this work. Instead, turbulence statistics are interpreted via the Kraichnan-Leith-Batchelor theory of 2D turbulence. In this framework, TKE values represent the energy stored in the large-scale coherent vortices. TKE production serves as a theoretical maximum for energy transfer from the mean shear, free from spanwise losses. Finally, the dissipation rate is interpreted as entropy dissipation, representing the viscous smoothing of vorticity gradients rather than kinetic energy loss. This framework establishes a 'best-case scenario' for vortex persistence. If macroscale structures are attenuated in this 2D limit, they would undoubtedly be destroyed in a fully 3D environment.

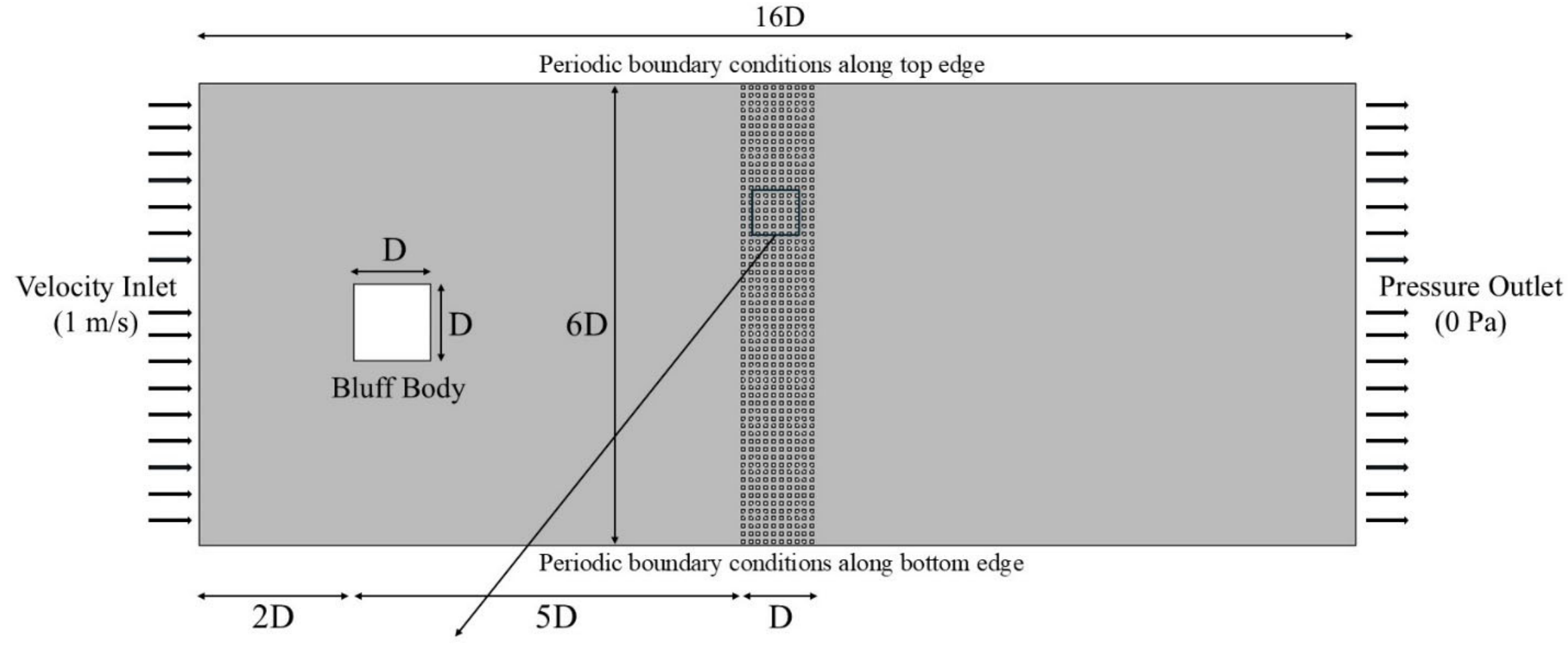

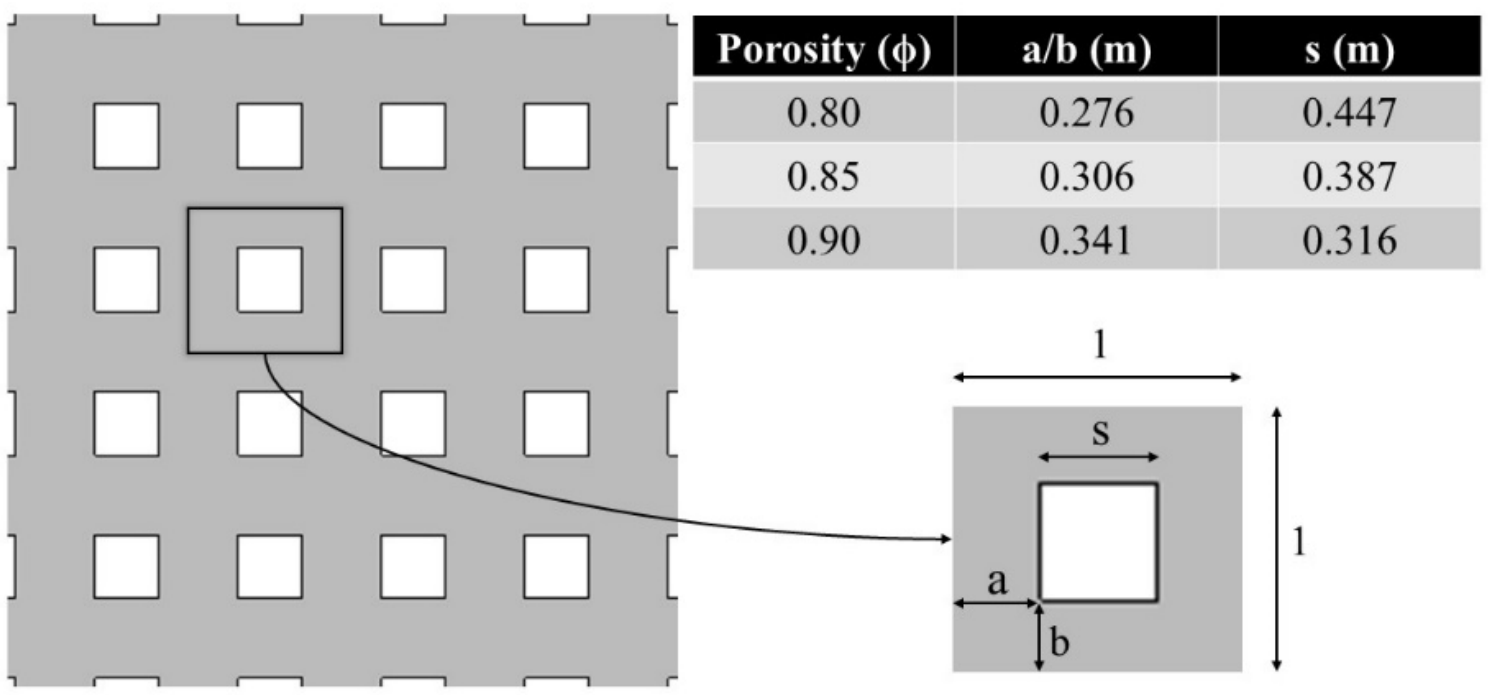

| Porosity (ϕ) | a/b (m) | s (m) |
|---|---|---|
| 0.80 | 0.276 | 0.447 |
| 0.85 | 0.306 | 0.387 |
| 0.90 | 0.341 | 0.316 |

**Fig. 1** The simulation geometry used to model macroscale vortex impingement on a porous layer.

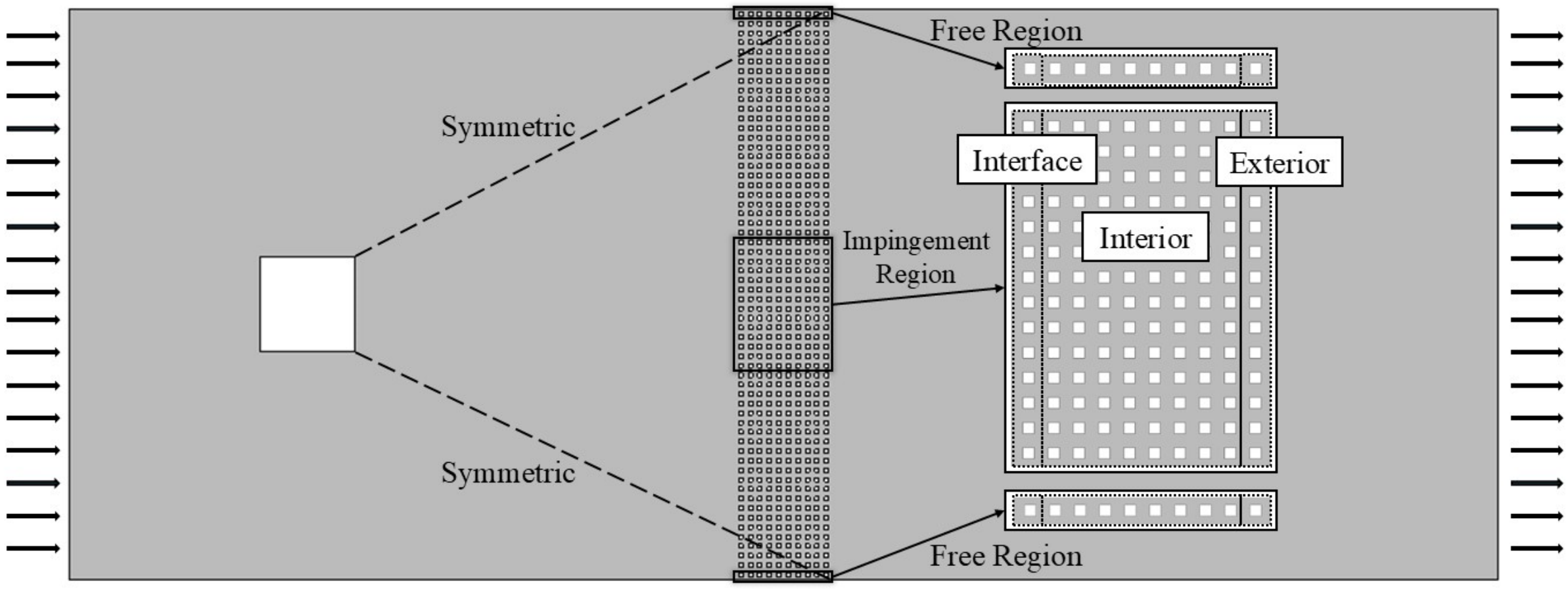

**Fig. 2** Region definitions used in the simulation (free-flow, impingement, interface, interior, and exterior).

**2.2 Numerical Method.** We solve the incompressible Navier–Stokes and energy equations using the Finite Volume Method in ANSYS Fluent 2025 R1. The governing equations of the flow are:

$$\frac{\partial u_j}{\partial x_j} = 0 \tag{3}$$

$$\frac{\partial \rho u_i}{\partial t} + \frac{\partial \rho u_i u_j}{\partial x_j} = -\frac{\partial p}{\partial x_i} + \frac{\partial}{\partial x_j}\left[\mu\left(\frac{\partial u_i}{\partial x_j} + \frac{\partial u_j}{\partial x_i}\right)\right] + \rho g_i \tag{4}$$

$$\frac{\partial \rho E}{\partial t} + \frac{\partial (\rho E + p) u_j}{\partial x_j} = \frac{\partial}{\partial x_j}\left[k\frac{\partial T}{\partial x_j}\right] + \frac{\partial}{\partial x_j}\left[u_j \mu\left(\frac{\partial u_i}{\partial x_j} + \frac{\partial u_j}{\partial x_i}\right)\right] \tag{5}$$

$$E = C_p(T - 298.15) + \frac{1}{2}{u_j}^2 \tag{6}$$

The pressure Poisson equation is solved along with the mass and momentum equations in a coupled manner. Spatial gradients are calculated using the Least-Squares Cell-Based method. Pressure is stored in a staggered grid with respect to the velocities. Convection terms for momentum and energy are discretized using QUICK to limit numerical diffusion in shear layers and thermal plumes. Time integration is performed with Second-Order Implicit advancement. The step size $\Delta t$ is chosen such that CFL $\leq 0.5$.

The conductive gradient at the solid boundary is used to obtain the local wall heat flux, and the mean Nusselt number is modeled using the following equation:

$$\overline{Nu} = \bar{q}" \frac{s}{k\,(T_w - T_{inlet})} \tag{7}$$

**2.3 Adequacy of the Grid Resolution.** We demonstrate that the grid resolution employed in the simulations accurately captures the hydrodynamic and thermal features of interest in this study. To limit computational cost, we assess grid adequacy for the two bounding porosities, $\phi = 0.80$ and $\phi = 0.95$. Since all simulations use the same systematic mesh design, the grid-resolution analysis for these two bounding cases is assumed to be representative of the intermediate porosities. The Reynolds number based on the characteristic length of the large solid obstacle is fixed at Re = 10000. The near-wall resolution is controlled such that the non-dimensional wall distance satisfies the following to ensure proper resolution of the viscous sublayer:

$$y^+ = \frac{y \rho u_\tau}{\mu} < 1 \tag{8}$$

To evaluate sensitivity to the overall grid density, we test three maximum grid spacings corresponding to $\Delta x/d$ = 0.01, 0.015, and 0.02, where $d$ is the obstacle spacing. Table 1 summarizes the percentage change in the total drag coefficient when the grid is refined. These differences remain small for $\phi = 0.80$, indicating satisfactory grid convergence. For $\phi = 0.95$, the more open geometry leads to stronger sensitivity, but the finest grid still yields acceptable accuracy for the purposes of this study.

**Table 1** Percentage change in the total drag coefficient due to grid refinement.

| Change in grid resolution | $\phi = 0.80$ | $\phi = 0.95$ |
|---|---|---|
| $\Delta x/d$ = 0.01–0.15 | 2.5% | 4.6% |
| $\Delta x/d$ = 0.15–0.20 | 11% | 11.8% |

**2.4 Terms of the Turbulence Kinetic Energy Transport Equations.** TKE transport is analyzed in this study by evaluating the dominant terms of production and dissipation rate. The production of TKE is defined as

$$P = -\sum_{i,j} \overline{u_i' u_j'} \frac{\partial \overline{u_i}}{\partial x_j}, \tag{9}$$

where $\overline{u_i}$ denotes the mean velocity, and $\overline{u_i' u_j'}$ represents the resolved Reynolds stresses obtained from time-averaged fluctuations. The TKE dissipation rate is evaluated as

$$\varepsilon = \nu \sum_{i,j} \overline{\frac{\partial u_i'}{\partial x_j} \frac{\partial u_i'}{\partial x_j}} \tag{10}$$

## 3. RESULTS AND DISCUSSION

**3.1 Large-scale coherent vortices are diminished at the porous/fluid interface.** The first objective of this paper is to establish the flow behaviour at the porous/fluid interface as turbulent vortices impinge on it to understand whether externally forced macroscale structures can penetrate the porous layer. We analysed the flow distribution across the entire simulation domain for the case of porosity = 0.80. The flow around the macroscale solid obstacle forms a classical von Kármán street behind the bluff body such that the generated vortices are turbulent in nature (figure 3), as is expected for Re = 10000. These vortices are roughly the size of the bluff body itself (order *D*), which establishes a clear separation between the macroscale turbulence generated by the obstacle and the microscale architecture of the pores within the porous layer. These macroscale vortices are advected by the flow and impinge on the porous layer. When these macroscale turbulent vortices impinge on the porous/fluid interface, we do not observe any macroscale turbulent structures persisting inside the porous layer, nor at the downstream locations behind the porous layer. This suggests that the macroscale vortex motion is completely reorganized inside the porous layer.

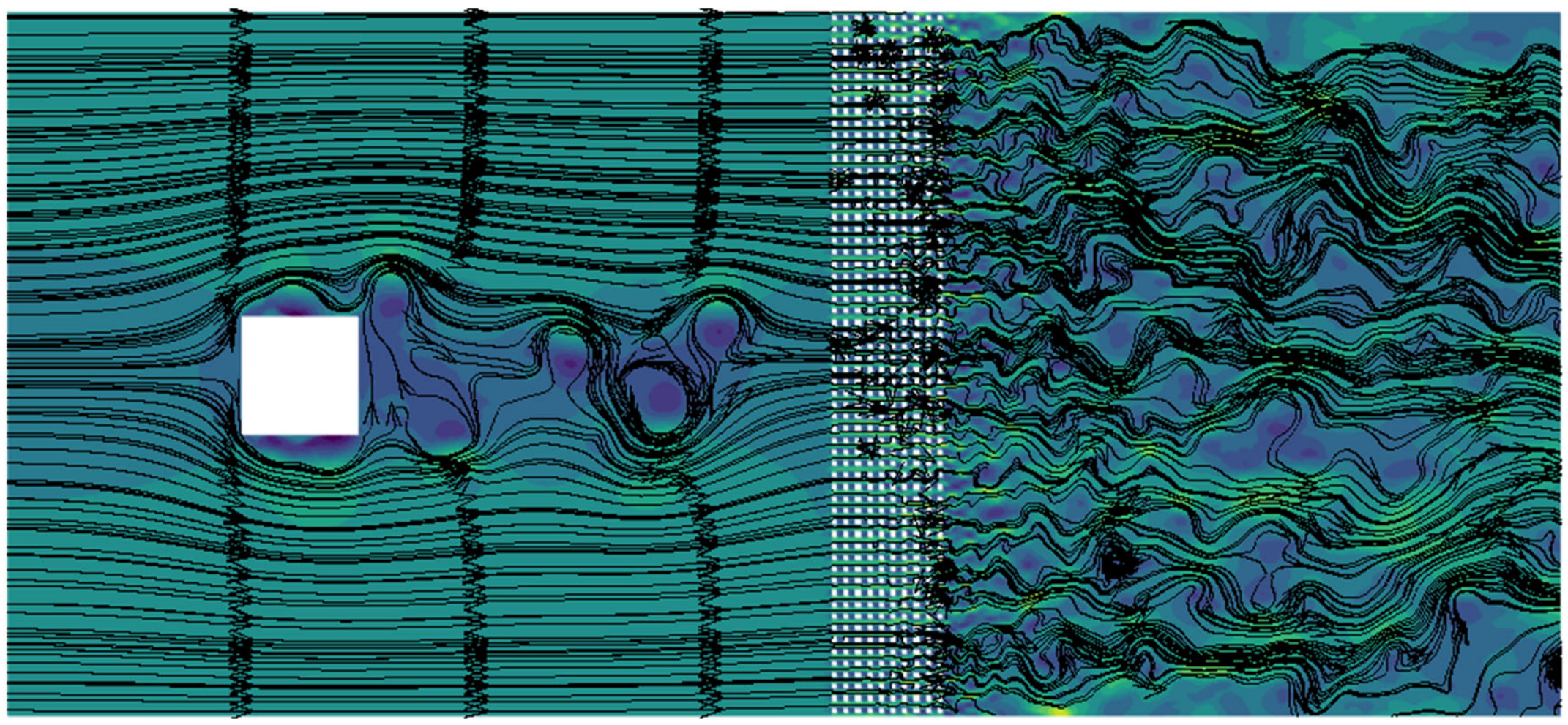

**Fig. 3** Contours of *x*-velocity overlaid with instantaneous flow streamlines.

The dissipation of the macroscale vortices is further evidenced by the vortex impingement over time at the porous/fluid interface (figure 4). At $t = t_0 + 0$ s, the macroscale vortex approaches the porous/fluid interface where it creates a localized region of low pressure and rotational flow streamlines. We then observed that the vortex diminishes in size rapidly upon impinging on the solid obstacles of the porous layer between $t = t_0 + 1$ s and $t = t_0 + 2$ s. Following this ($t = t_0 + 3$ s), there is no evidence of the rotational flow streamlines, a characteristic of the macroscale vortex motion, persisting within the pores. However, we observe that the momentum deficit of the vortex wake region creates localized regions of low flow velocity. This implies that even though the macroscale vortex cannot enter the porous layer, the dynamic motion of the vortex street persists inside the porous layer in the region of impingement.

We also observed that the flow streamline tortuosity immediately following the impingement of the macroscale vortex is increased, when compared to the regions that are not influenced by flow impingement. Streamlines are not confined to exist along an in-line array of porous obstacles but can be transported between adjacent rows due to the highly chaotic nature of the flow. The exact mechanism of this chaotic flow is not completely understood. We observed that the porous layer begins to produce vortical motions behind the individual solid obstacles leading to production of microscale coherent structures. Secondary flow instabilities induced by this vortex production are known to enhance oscillatory behavior of the vortex wake in these intermediate porosity values, resulting in tortuous flow streamlines. However, the lack of chaotic flow behavior away from the impingement region contradicts this explanation. We hypothesize that tortuous flow streamlines are only induced when two complementary mechanisms are present: the secondary flow instability and the laminar-turbulent transition of the flow. In the absence of laminar-turbulent transition, the regularity of the flow patterns makes it difficult for the vertical transport between rows of solid obstacles in the porous layer.

We confirmed that this macroscale vortex breakdown at the porous/fluid interface occurs for all the tested values of porosity, 0.80-0.95 (section 3.3). The mechanism of this vortex breakdown is determined by turbulence transport inside the porous layer, specifically the turbulence kinetic energy, and the production and dissipation rate of turbulence. The distribution of turbulence kinetic energy (figure 5) mirrors our intuitive observations of vortex dissipation at the porous/fluid interface. A "hot band" of turbulence kinetic energy is observed at the region of vortex impingement, which is immediately followed by a high magnitude of turbulence kinetic energy around the first column of solid obstacles. At this location on the interface, we observed pronounced shear due to the formation of high-velocity channels aligned with pore passages between the solid obstacles. This region of high turbulence kinetic energy is immediately diminished, starting in the second column of solid obstacles and creating a region of low turbulence kinetic energy following the porous/fluid interface. As microscale vortices are produced by the individual solid obstacles of the porous medium, we observe that the turbulence kinetic energy begins to increase as more turbulence is continually produced by the porous medium itself.

Analyzing the turbulence kinetic energy budget through the production and dissipation rate of TKE provides a physical explanation for these observed trends (figure 6). The porous/fluid interface is characterized by negative turbulence production at the forward faces of the solid obstacles as well as a high dissipation rate around the solid obstacle surface where the flow experiences shear. While negative turbulence production is a rare occurrence in other canonical turbulent flows, it is frequently observed in turbulent flow in porous media due to the presence of vortical flow and flow streamline tortuosity that create localized negative strain rate [11]. These factors combined are very conducive to the destruction of macroscale coherent structures, resulting in the “hot band” of TKE at the impingement region that does not extend into the porous layer. Following this destruction of TKE at the interface, we observe the production of TKE deeper inside the porous layer caused by the formation of shear layers around the solid obstacle surface because of vortex formation. The volumetric production of TKE in the unit cell inside the porous layer outweighs the volumetric dissipation rate, resulting in a net TKE deep inside the porous layer at the microscale level. However, since this turbulence is generated locally at the microscales by the porous medium rather than from the macroscale vortex wake, we conclude that macroscale turbulence cannot survive inside a porous layer under these flow conditions.

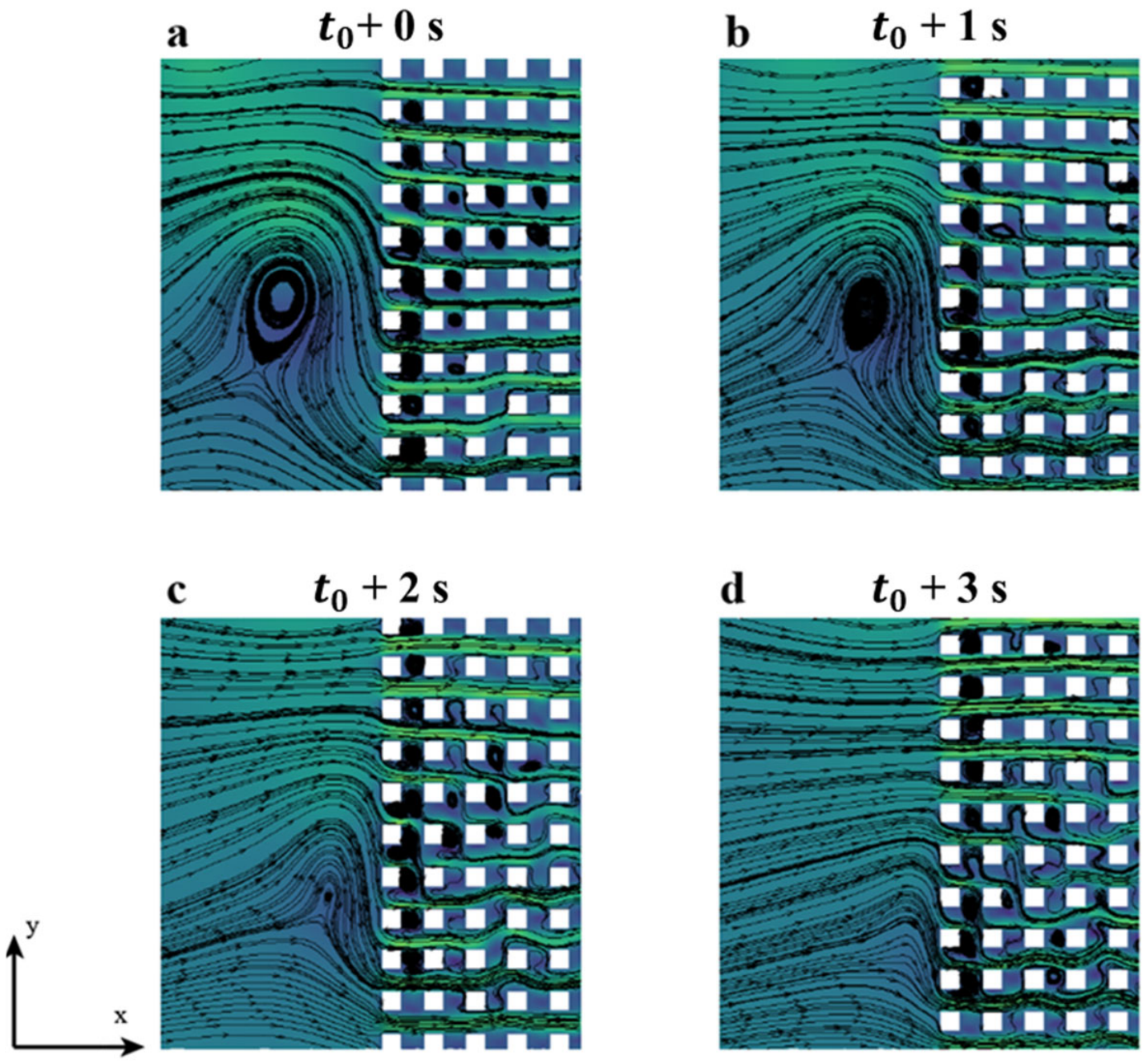

**Fig. 4** Contours of x-velocity and streamlines at successive times for a porosity of 0.80 after statistical steady state is reached. The first panel is shown at the reference time $t_0 = 3850$ s, while the remaining panels are shown at 1 s intervals thereafter.

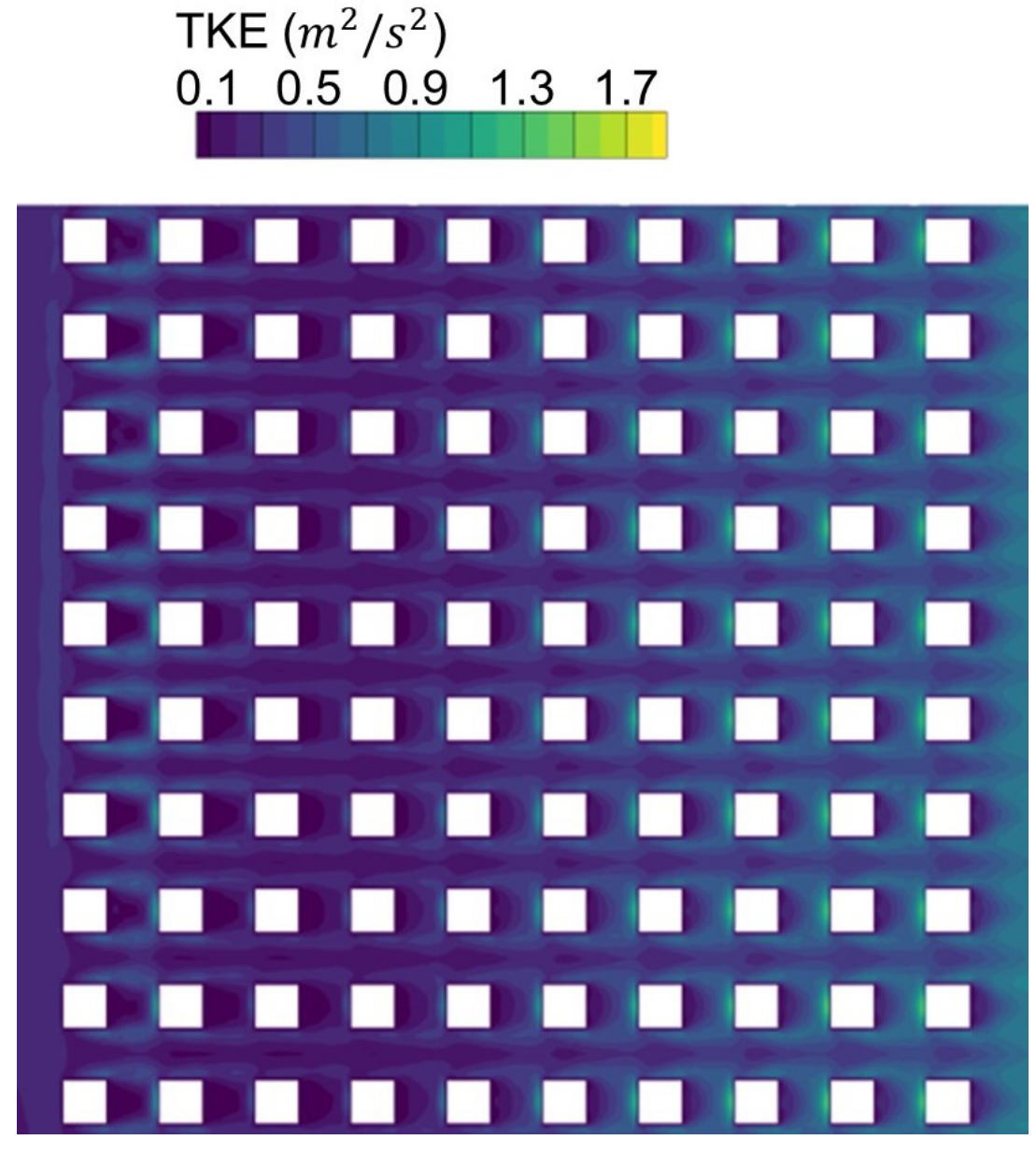

**Fig. 5** TKE drops at the porous/fluid interface and reappears as smaller-scale streaks within the first few pores.

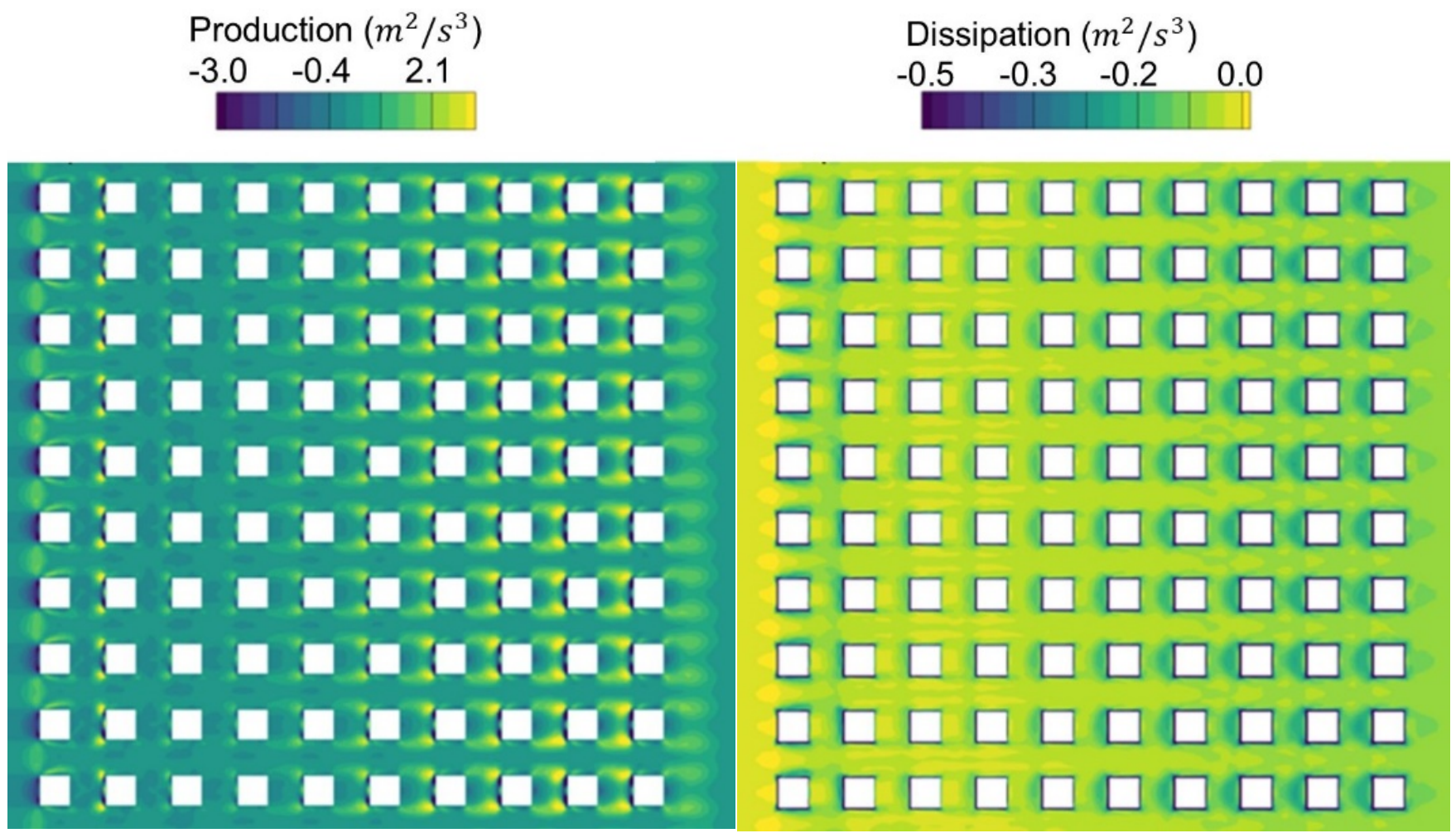

**Fig. 6** Shear production (left) and dissipation hotspots (right) concentrate near upstream obstacle faces and pore entrances, forming interfacial bands that persist into downstream rows.

**3.2 Heat-Transfer Intensification at Porous Interfaces: Linking Shear, TKE, and Nusselt number patterns.** The destruction of macroscale vortices and the subsequent production of microscale vortices have strong implications for the heat transfer characteristics in the impingement region. Heat transfer in this region is determined by two competing factors: (i) the macroscale vortices will induce strong mixing at the interface and accelerate transition to turbulence deep inside the porous layer, (ii) the momentum deficit created by the vortex shedding will decrease the local flow velocity and increase boundary layer thickness. We analysed the temperature distribution inside the porous layer and the Nusselt number distributions at four regions of interest, namely the interface and interior locations in the vortex impingement and free-flow regions.

The distribution of mean static temperature shows that the temperature of the fluid increases along the depth of the porous layer as the fluid encounters more solid obstacles held at a higher temperature (figure 7). While this increase in bulk temperature of the fluid is expected, it does not occur in a linear fashion. The thinnest momentum and thermal boundary layers are observed at the porous/fluid interface where strong local shear is generated by the channeling of the flow between the solid obstacles. In the solid obstacles immediately following the interface, the macroscale vortex has been broken down, and it is followed by a weakly turbulent flow region where the vortices that are formed behind the solid obstacles are recirculating in nature. These recirculating vortices hinder heat transfer at the solid obstacle surface and trap heat in the horizontal void space between the solid obstacles. As microscale vortices develop deep inside the porous layer, the vortices experience a secondary flow instability that oscillates the vortex wake around the solid obstacles, similar to observations in previous work [12,13]. The vortices become shedding vortices that enable greater heat transfer and a more diffuse temperature distribution that is characteristic of turbulent mixing flows [14].

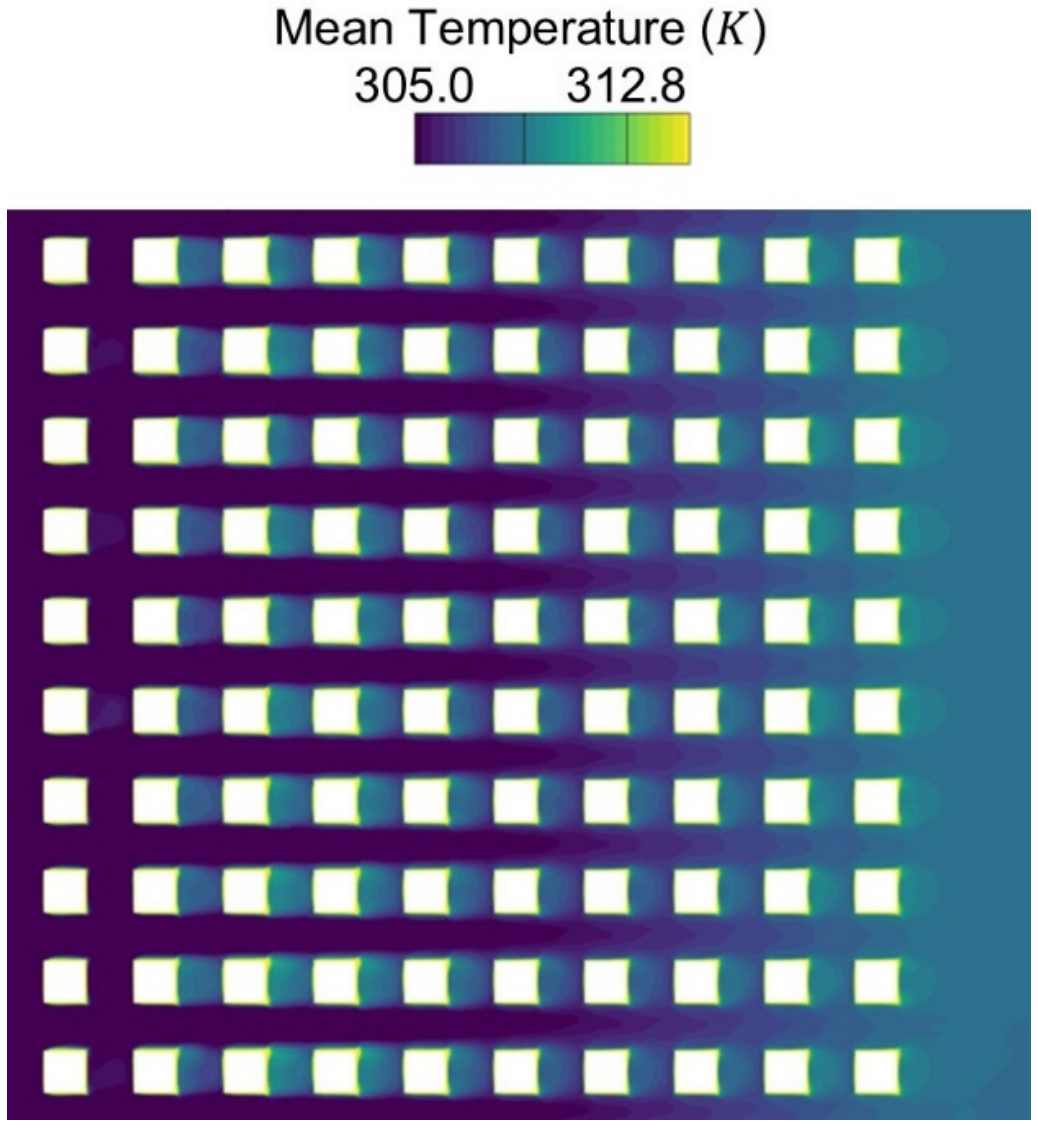

**Fig. 7** Mean static temperature (K) shown at the same magnified view of the porous layer.

Consequently, we observe that the Nusselt number on the solid obstacle surface assumes its maximum value at the forward-facing vertex of the solid obstacle geometry (figures 8 and 9), which is the stagnation point of the flow. Thus, the temperature gradient between the cold fluid entering the pore and the hot solid obstacle is at the highest value, and the boundary layer thickness is the smallest, resulting in a higher surface-averaged Nusselt number in these regions. This peak in Nusselt number is followed by a rapid decrease to a uniform value, indicating that the thermal boundary layer develops rapidly along the obstacle surface. The peak value of the Nusselt number is the highest at the porous/fluid interface in both the impingement and free-flow regions. However, we observed that the peak Nusselt number in the vortex impingement region is consistently smaller than the peak Nusselt number in the free-flow region. Remarkably, these differences in the surface Nusselt number between impingement and free-flow regions are not significant in the interior of the porous layer.

Taken together, these results show that both the impingement zone and the free-stream region exhibit intense but highly localized Nusselt-number hot spots. Heat transfer is highest at the porous/fluid interface; however, the impingement region shows a reduced heat flux relative to the free-stream region because the wake momentum deficit lowers the local near-interface velocity. Within the porous layer, heat transfer is generally lower than at the interface, but it increases with depth as microscale vortices form and strengthen around the solid obstacles. This increase is consistent with enhanced mixing associated with the onset of turbulence and secondary flow instabilities deeper in the porous matrix.

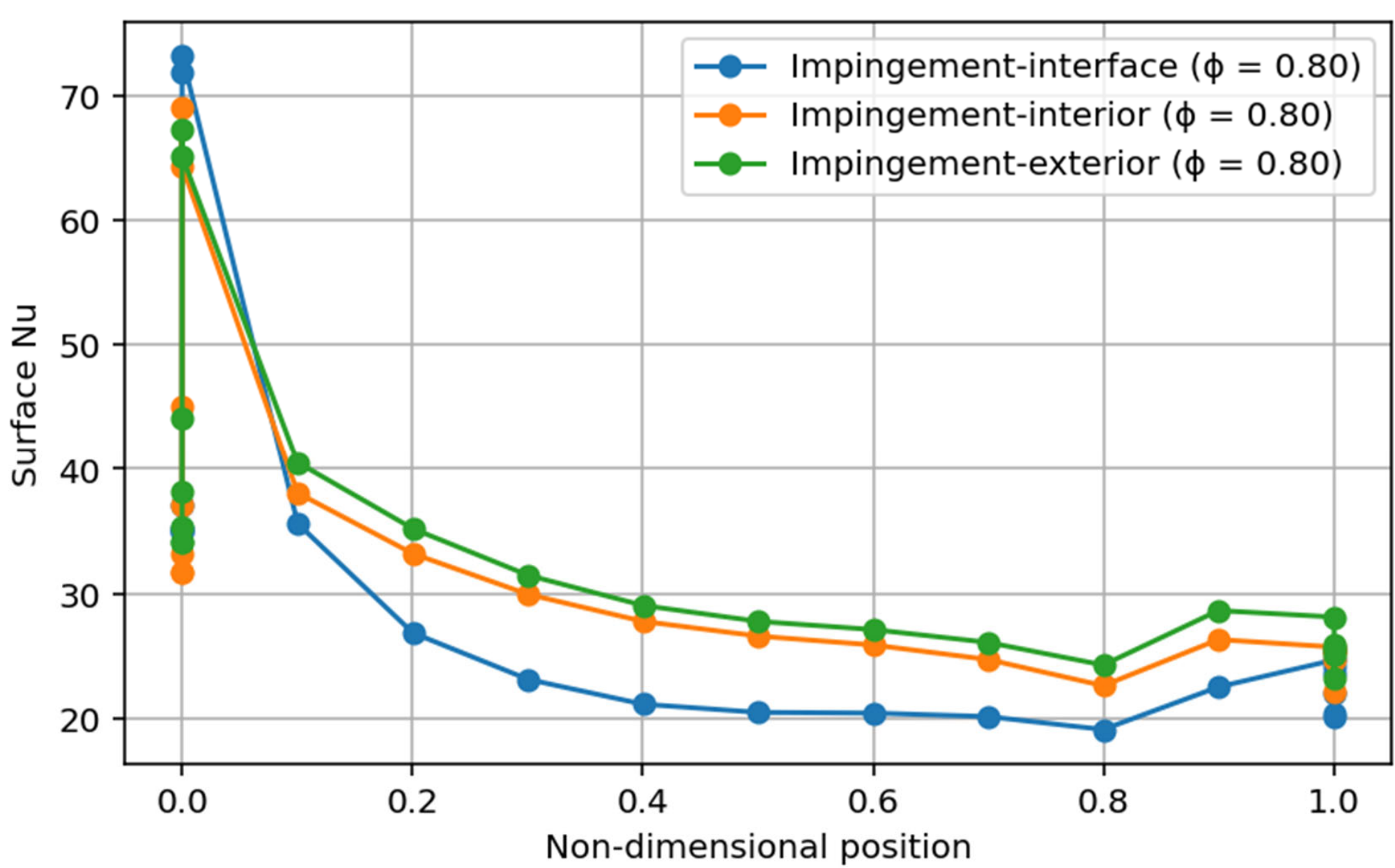

**Fig. 8** Time-averaged surface Nusselt number along the interface, interior, and exterior obstacles located in the impingement region ($\phi = 0.80$).

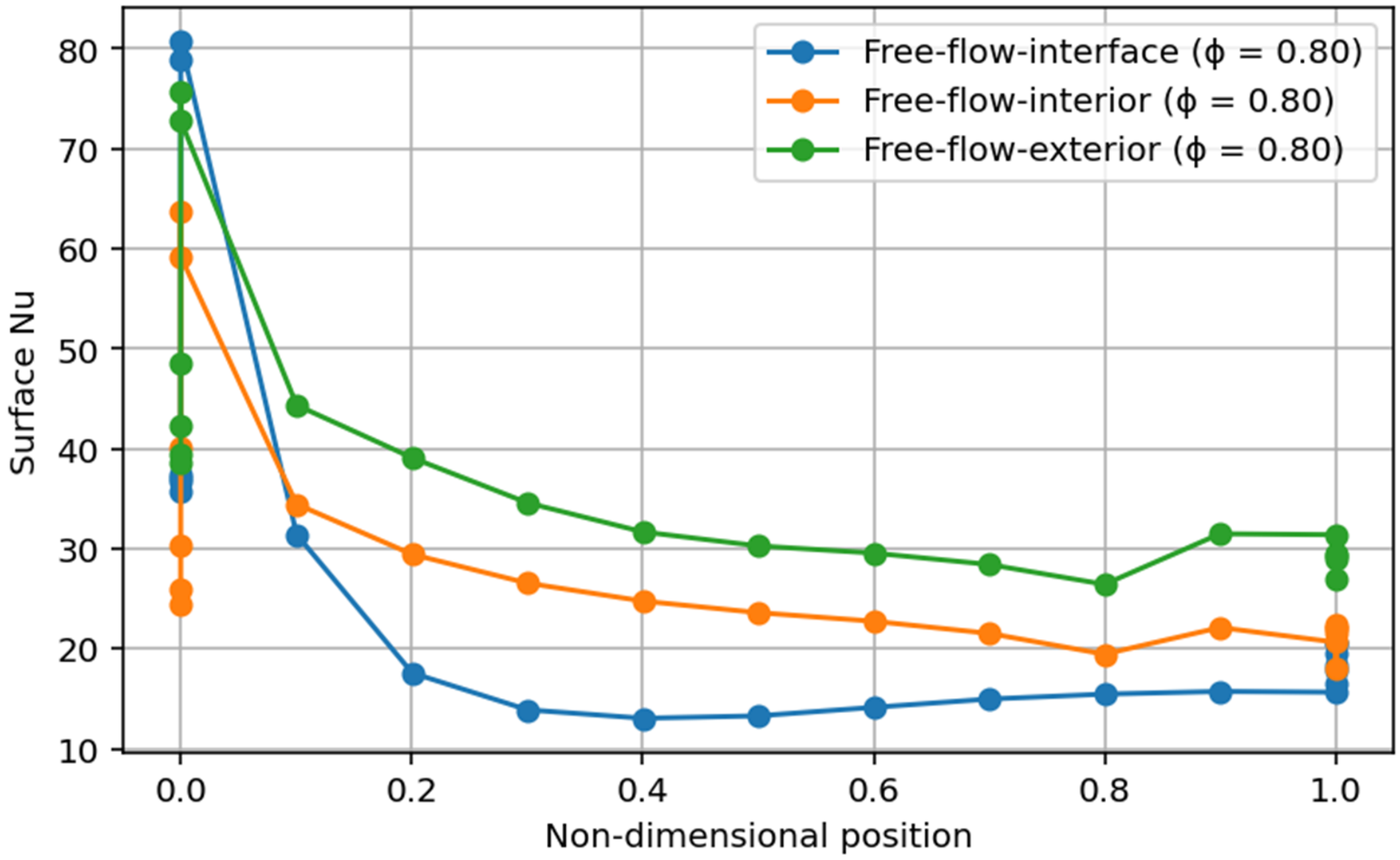

**Fig. 9** Time-averaged surface Nusselt number for top obstacles in the interface, interior, and exterior columns within the free-stream region ($\phi = 0.80$).

**3.3 Effect of Porosity on Turbulence Dissipation and Surface Heat Transfer.** Porosity plays a crucial role in the vortex dynamics at the porous/fluid interface and how the flow subsequently develops inside the porous layer. To analyze the flow dependence on porosity, we computed the volume-averaged distributions of mean static temperature, production, dissipation rate and TKE over the entire porous layer and compared the results for $\phi = 0.80$ and 0.95 (figure 10). These distributions show a clear distinction in the flow properties between the impingement and free-flow regions. We observe the formation of a temperature hot spot with a higher volume-averaged temperature in the impingement region when compared to the free-flow region. This observation is directly explained by the low TKE in the region of the hot spot. While the impingement of the macroscale vortex introduces TKE at the porous/fluid interface, the momentum deficit created by the vortex delays the production of TKE. This results in lower TKE deep inside the porous layer in the impingement region when compared to the same location in the free-flow region.

This observation is consistent across both $\phi = 0.80$ and 0.95; however, the temperature hot spot has a greater magnitude at low porosity ($\phi = 0.80$) when compared to high porosity ($\phi = 0.95$). At higher porosity ($\phi = 0.95$), the production of TKE approaches its peak value earlier in the porous layer than at lower porosity. This is caused by the enhanced dynamics of the microscale flow at high porosity due to the increased pore space in between the solid obstacles. At lower porosity ($\phi = 0.80$), the more constricted flow yields stronger shear, higher TKE and dissipation, but this transition is delayed by the restriction of the vortex dynamics imposed by the smaller pores. Thus, the steep temperature increase along the depth of the porous layer at low $\phi$ results from the combined effects of higher local turbulence intensity, which enhances mixing, and a larger surface-area-to-volume ratio, which strengthens fluid/solid thermal interaction. This large surface-area-to-volume ratio at low porosity leads to increased surface Nusselt numbers when compared to high porosity (figure 11), as well as a larger peak value of Nusselt number value due to higher flow stagnation pressure. Consistent with this behavior, figures 12 and 13 show that the lower-porosity case ($\phi = 0.80$) exhibits higher local, surface-averaged, and time-averaged Nusselt numbers at all representative surface locations, as well as systematically higher mean temperatures within the porous region. In contrast, the higher-porosity case ($\phi = 0.95$) displays reduced heat-transfer levels despite similar macroscale dissipation trends.

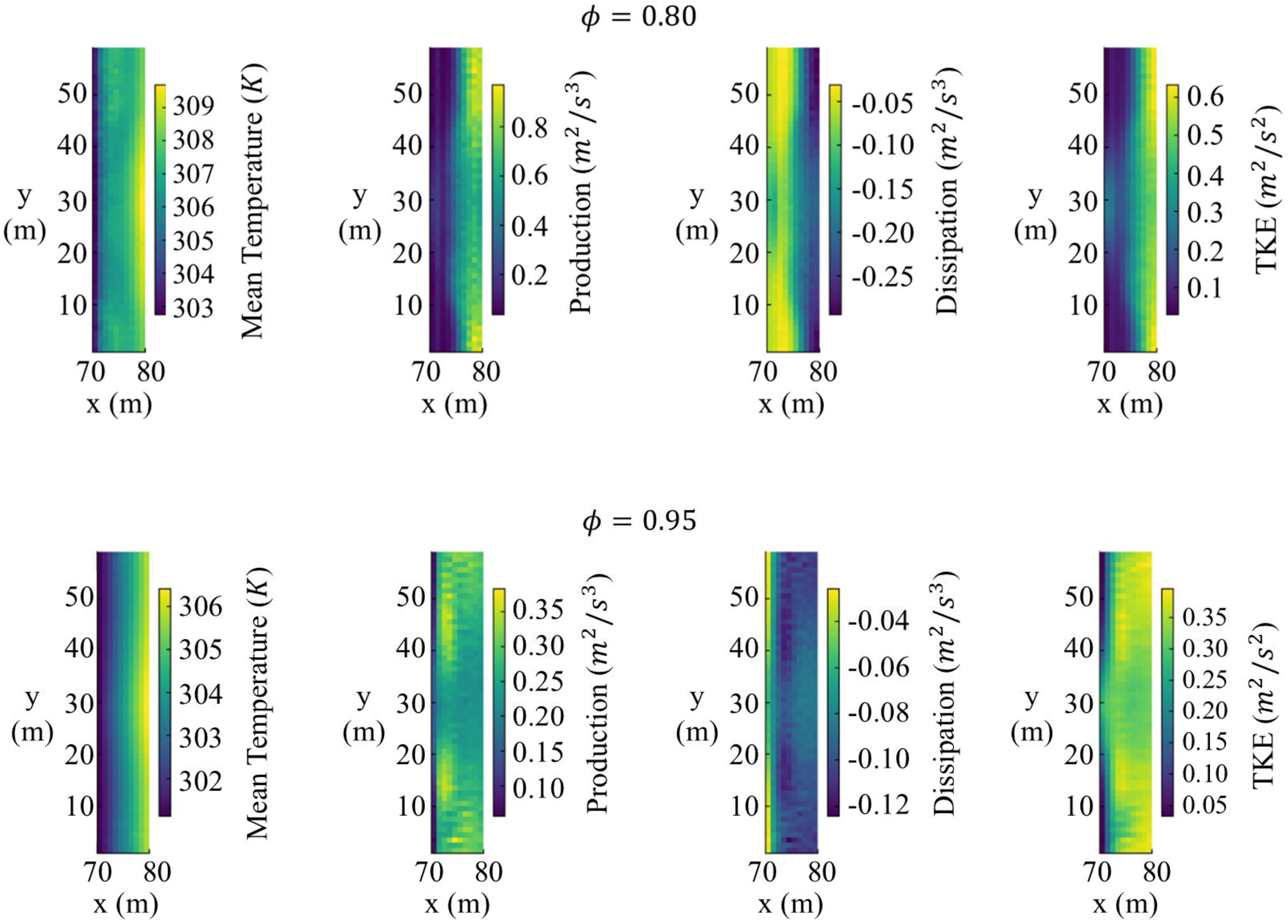

**Fig. 10** Effect of porosity on turbulence production, dissipation, TKE, and temperature in the porous layer.

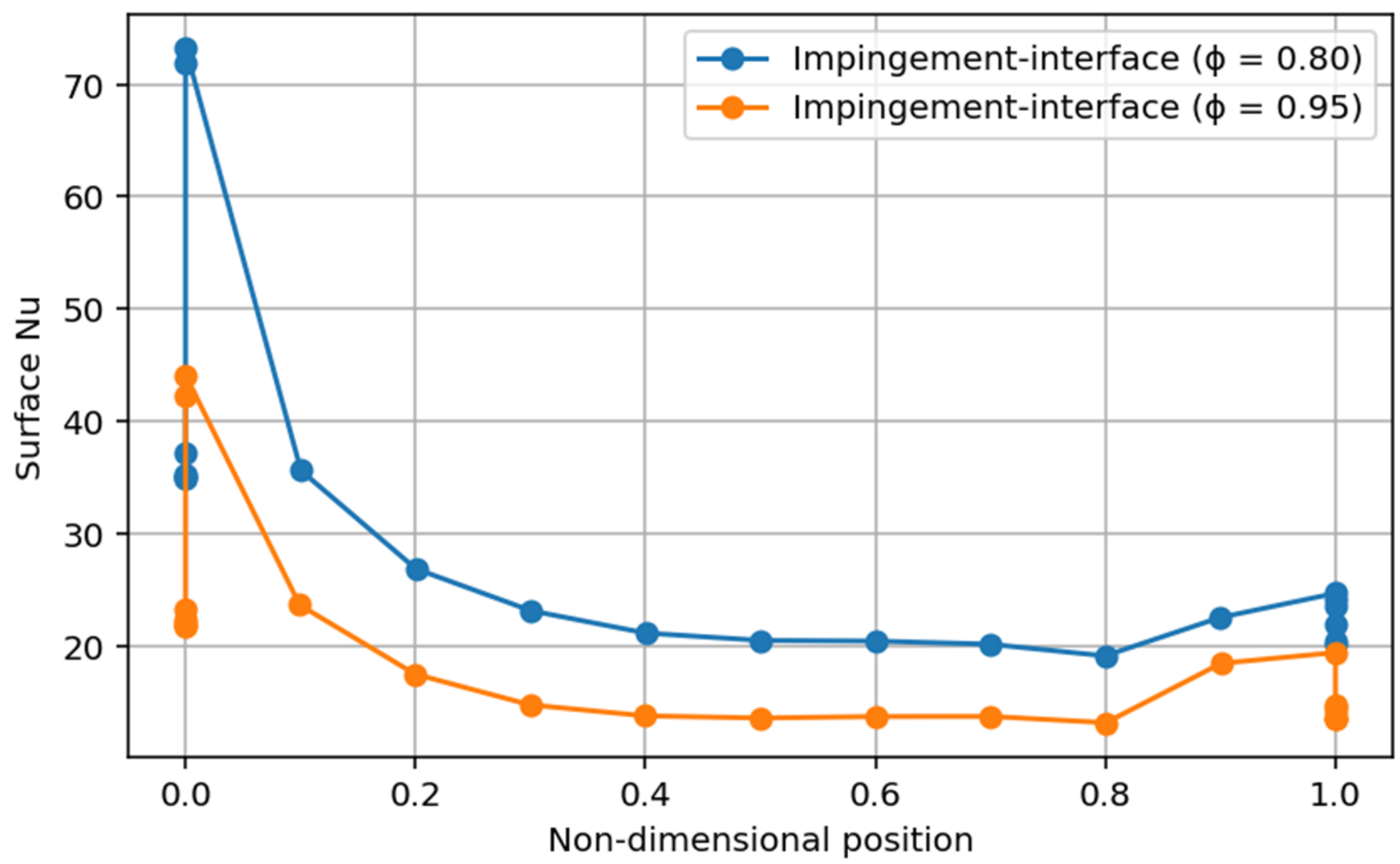

**Fig. 11** Time-averaged surface Nusselt number along the porous/fluid interface row for two porosities, $\phi$ = 0.80 and $\phi$ = 0.95.

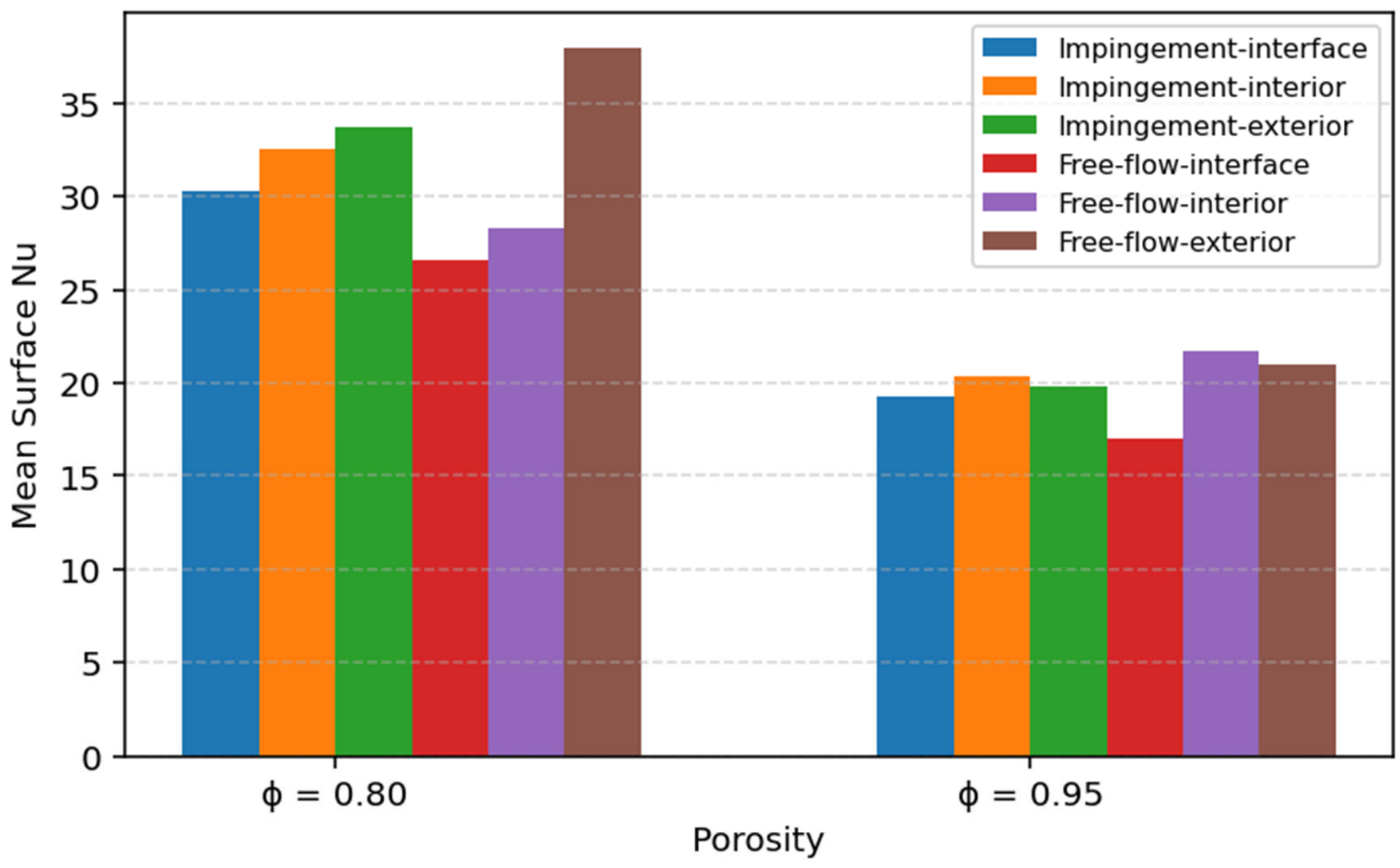

**Fig. 12** Average mean surface Nusselt number at six locations within the porous configuration (impingement-interface, impingement-interior, impingement-exterior, free-interface, free-interior, and free-exterior) for two porosities, $\phi$ = 0.80 and $\phi$ = 0.95.

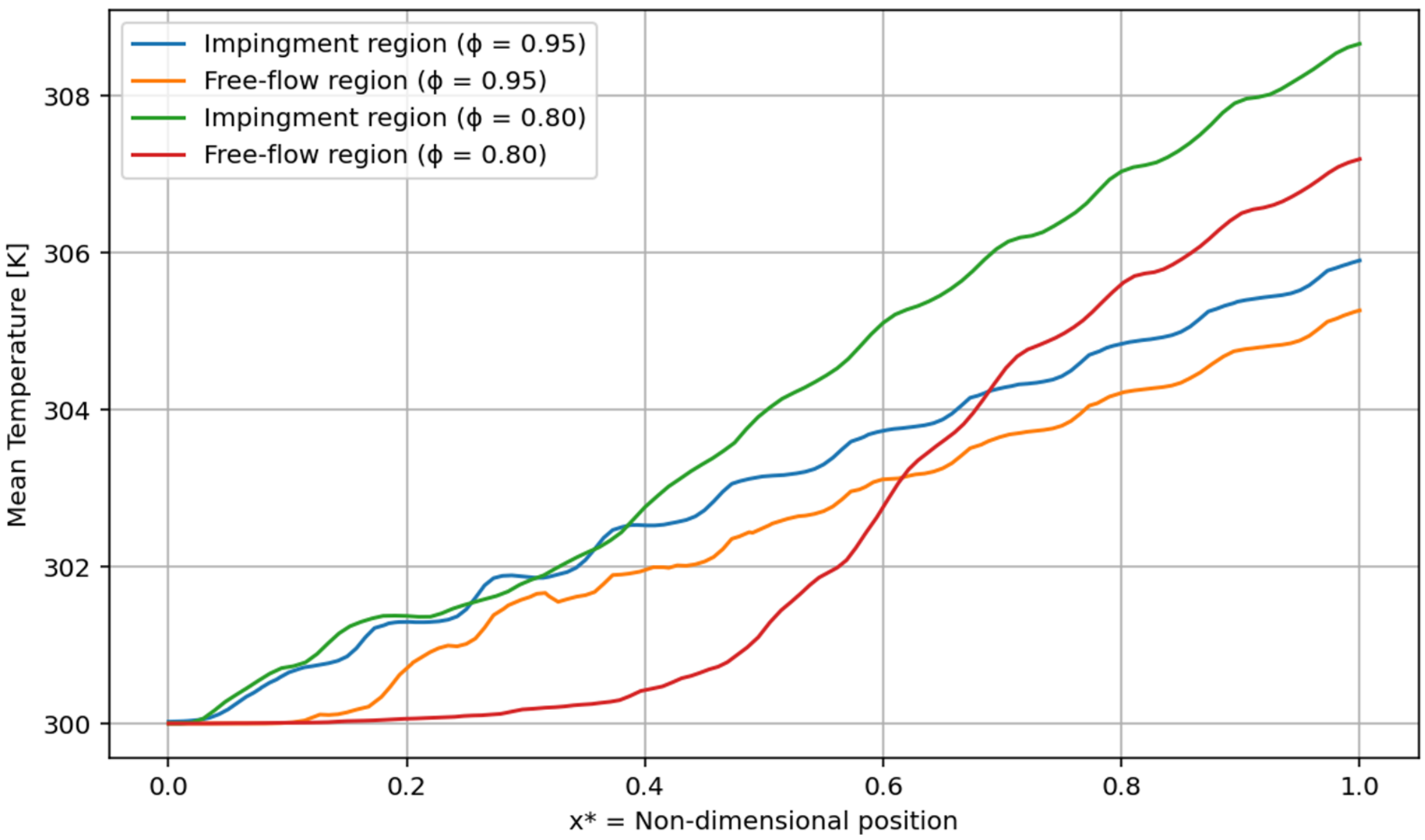

**Fig. 13** Mean temperature distribution in the porous layer as a function of the normalized streamwise coordinate $x^*$, where $x^* = 0$ corresponds to the porous/fluid interface and $x^* = 1$ denotes the exterior of the porous layer, for two transverse locations (in the impingement region and in the free-flow region) and two porosities ($\phi = 0.80$ and $\phi = 0.95$).

## 4. CONCLUSION

This study utilized 2D Direct Numerical Simulation to demonstrate that macroscale vortices are unable to penetrate a microscale porous layer, addressing the question of whether macroscale turbulence can survive inside porous media. We observed that the incoming wake structures undergo immediate breakdown at the porous/fluid interface, creating a localized "hot band" of elevated turbulence kinetic energy (TKE) characterized by intense shear production and viscous dissipation. This behavior was consistent for the porosities investigated here ($\phi = 0.80$ and $\phi = 0.95$), representing low-porosity and high-porosity media, respectively. Consequently, the turbulence observed deeper within the porous matrix is not a direct remnant of the externally forced vortex wake; rather, it is regenerated locally through microscale vortex shedding and recirculation behind individual solid obstacles in the porous layer.

This confirms that the porous interface primarily acts as a spectral filter that converts macroscale wake energy into pore-scale motions before the flow propagates inward. Under these conditions, thermal performance is determined by the interplay between interfacial vortex breakdown and the depth of flow penetration. A lower porosity ($\phi = 0.80$) enhances flow confinement and increases local shear near the interface. This results in higher Nusselt numbers at the interface and in most porous subregions, effectively concentrating heat transfer near the exposed surfaces. Conversely, higher porosity ($\phi = 0.95$) reduces peak interfacial heat flux and shifts the location of the maximum mean Nusselt number toward interior/free-flow-region sites (free-flow-interior), reflecting a redistribution of near-interface momentum pathways in the more open matrix. These findings establish porosity as a tunable design parameter for porous-coated heat exchangers, enabling trade-offs between localized interfacial cooling and more distributed thermal interaction within the porous layer.

## NOMENCLATURE

| | | |
|---|---|---|
| $a, b$ | pore opening side lengths in unit cell | (m) |
| $a_s$ | wetted perimeter per unit area | ($\mathrm{m^{-1}}$) |
| $D$ | bluff-body side length / reference outer scale | (m) |
| $D_h$ | hydraulic diameter of porous surface | (m) |
| $E$ | specific total energy | ($\mathrm{J\ kg^{-1}}$) |
| $f$ | shedding frequency | ($\mathrm{s^{-1}}$) |
| $k$ | thermal conductivity | ($\mathrm{W\ m^{-1}\ K^{-1}}$) |
| $L$ | characteristic length used in Nu | (m) |
| $L_p$ | penetration length inside porous matrix | (m) |
| $p$ | static pressure | (Pa) |
| $q"$ | local wall heat flux | ($\mathrm{W\ m^{-2}}$) |
| $s$ | streamwise pitch of porous lattice | (m) |
| $T$ | temperature | (K) |
| $T_b$ | upstream bulk temperature at porous layer entrance | (K) |
| $T_\infty$ | inflow/surrounding (reference) temperature | (K) |
| $T_w$ | isothermal solid (porous matrix) temperature | (K) |
| $u$ | velocity vector/components | ($\mathrm{m\ s^{-1}}$) |
| $U_\infty$ | uniform inlet velocity | ($\mathrm{m\ s^{-1}}$) |
| $\Delta p$ | pressure drop across porous layer | (Pa) |